\documentclass{elsart}

 \usepackage{graphicx}

\usepackage{amssymb}

\journal{Physica A}

\begin{document}

\begin{frontmatter}



\title{The random field Ising model with an asymmetric trimodal probability distribution}


\author{I. A. Hadjiagapiou\corauthref{cor1}}
\corauth[cor1]{Corresponding author.}
\ead{ihatziag@phys.uoa.gr}

\address{Section of Solid State Physics, Department of Physics,
University of Athens, Panepistimiopolis, GR 15784 Zografos,
Athens, Greece}

\begin{abstract}

The Ising model in the presence of a random field is investigated
within the mean field approximation based on Landau expansion. The
random field is drawn from the  trimodal probability distribution
$P(h_{i})=p \delta(h_{i}-h_{0}) + q \delta (h_{i}+h_{0}) + r
\delta(h_{i})$, where the probabilities $p, q, r$ take on values
within the interval $[0,1]$ consistent with the constraint
$p+q+r=1$ (asymmetric distribution), $h_{i}$ is the random field
variable and $h_{0}$ the respective strength. This probability
distribution is an extension of the bimodal one allowing for the
existence in the lattice of non magnetic particles or vacant
sites. The current random field Ising system displays second order
phase transitions, which, for some values of $p, q$ and $h_{0}$,
are followed by first order phase transitions, thus confirming the
existence of a tricritical point and in some cases two tricritical
points. Also, reentrance can be seen for appropriate ranges of the
aforementioned variables. Using the variational principle, we
determine the equilibrium equation for magnetization, solve it for
both transitions and at the tricritical point in order to
determine the magnetization profile with respect to $h_{0}$.

\end{abstract}
\date{\today}

\end{frontmatter}

\newpage
\section{Introduction}

The pure models for crystalline materials can describe the
respective experimental samples in exceptional situations, since
such a sample can contain impurities, broken bonds, defects, etc.,
making real physical systems to never be translationally
invariant, thus necessitating the modification of pure models
appropriately for comparing the experimental results with the
theoretical predictions. A small amount of quenched randomness can
influence significantly the phase transitions replacing a
first-order phase transition (FOPT) by a second-order phase
transition (SOPT), so that tricritical points and
critical-end-points are suppressed \cite{huiberker}. In two
dimensions, an infinitesimal amount of field randomness destroys
any FOPT. One such situation is the presence of random magnetic
fields acting on each spin in an otherwise free of defects
lattice; the respective pure system is considered to be described
by an Ising model, which is now transformed into the random field
Ising model (RFIM) in the presence of random fields
\cite{physicstoday,natermannvillain,imryma}. RFIM had been the
standard vehicle for studying the effects of quenched randomness
on phase diagrams and critical properties of lattice spin-systems
and had been studied for many years since the seminal work of Imry
and Ma \cite{imryma}. Associated with this model are the notions
of lower critical dimension, tricritical points, higher order
critical points and random-field probability distribution function
(PDF). The simplest model exhibiting a tricritical phase diagram
in the absence of randomness is the Blume-Capel model -- a regular
Ising spin-$1$ model \cite{blume,capel}. Although much effort has
been invested for the study of the RFIM, the only well-established
conclusion is the existence of a phase transition for $d \geq 3$
(d space dimension), that is, the critical lower dimension $d_{l}$
is 2 after a long controversial discussion \cite{imryma,imbrie},
while many other issues are still unanswered; among them is the
order of the phase transition, the existence of a tricritical
point (TCP) and the dependence of these on the form of the random
field PDF. According to the mean-field approximation (MFA) the
choice of the random field PDF can lead to a continuous
ferromagnetic/paramagnetic (FM/PM) boundary as in the single
Gaussian, whereas for the bimodal this boundary can be divided
into two parts, an SOPT branch for high temperatures and an FOPT
branch for low temperatures separated by a TCP at
$kT^{t}_{c}/(zJ)=2/3$ and $h^{t}_{c}/(zJ)=(kT^{t}_{c}/(zJ)) \times
\arg\tanh(1/\sqrt{3})\simeq 0.439$
\cite{aharony,sneiderpytte,fernandez}, where $z$ is the
coordination number, $k$ is the Boltzmann constant and $T^{t}_{c},
h^{t}_{c}$ are the tricritical temperature and random field,
respectively, such that for $T<T^{t}_{c}$ and $h>h^{t}_{c}$ the
transition to the FM phase is of first order. However, this
behavior is not fully elucidated since in the case of the three
dimensional RFIM, the high temperature series expansions yield
only continuous transitions for both PDF's \cite{gofman};
according to Houghton et al \cite{houghton} both distributions
predict a tricritical point with $h^{t}_{c} = 0.28 \pm 0.01$ and
$T^{t}_{c} = 0.49\pm 0.03$ for the bimodal and $\sigma^{t}_{c} =
0.36 \pm 0.01$ and $T^{t}_{c} = 0.36\pm 0.04$ for the Gaussian
with critical standard deviation $\sigma^{t}_{c}$. Galam and
Birman studied the crucial issue for the existence of a TCP within
the mean-field theory for a general PDF $p(\overrightarrow{H})$ by
using an even-degree free energy expansion up to eighth degree in
the order parameter; they proposed some inequalities between the
derivatives of the PDF up to sixth order at zero magnetic field
for the possible existence of a TCP \cite{galambirman2}. In Monte
Carlo studies for $d = 3$, Machta et al \cite{machta}, using the
Gaussian distribution, could not reach a definite conclusion
concerning the nature of the transition, since for some
realizations of randomness the magnetization histogram was
two-peaked (implying an SOPT) whereas for other ones was
three-peaked implying an FOPT; Middleton and Fisher
\cite{middleton}, using a similar distribution for $T = 0$,
suggested an SOPT with a small order-parameter exponent $\beta =
0.017(5)$; Fytas et al \cite{fytas1}, following Wang-Landau and
Lee entropic sampling schemes for the bimodal distribution
function with $h_{0} = 2$ and $h_{0} = 2.25$ for a simple cubic
lattice, concluded that their results indicated an SOPT by
applying the Lee-Kosterlitz free energy barrier method;
Hern$\acute{a}$ndez and coworkers claim they have found a
crossover between an SOPT and an FOPT at a finite temperature and
magnetic field for the bimodal distribution function
\cite{hernandezetal}. One of the main issues was the experimental
realization of random fields. Fishman and Aharony \cite{fishaha}
showed that the randomly quenched exchange interactions Ising
antiferromagnet in a uniform field $H$ is equivalent to a
ferromagnet in a random field with the strength of the random
field linearly proportional to the induced magnetization. Also
another interesting result found by Galam \cite{galam3} via MFA
was that the Ising antiferromagnets in a uniform field with either
a general random site exchange or site dilution have the same
multicritical space as the random-field Ising model with bimodal
PDF.

The usual PDF for the random field is either the symmetric bimodal

\begin{equation}
 P(h_{i})=p\delta(h_{i}-h_{0}) + q \delta (h_{i}+h_{0})  \label{bimodalp}
\end{equation}

where $p$ is the fraction of lattice sites having a magnetic field
$h_{0}$, while the rest fraction has a field $(-h_{0})$ and $p = q
= \frac{1}{2}$ \cite{aharony,andelman1,kaufkan}, or the Gaussian,
single or double symmetric,

\smallskip

\begin{eqnarray}
P(h_{i}) & = & \frac{1}{(2 \pi \sigma ^{2})^{1/2}} \; exp\left[-
\frac{h^{2}_{i}}{2 \sigma ^{2}}\right]             \nonumber  \\
P(h_{i}) & = & \frac{1}{2}  \frac{1}{(2 \pi \sigma^{2})^{1/2}}
\left\{exp\left[-\frac{(h_{i}-h_{0})^{2}}{2 \sigma^{2}}\right] +
exp\left[-\frac{(h_{i}+h_{0})^{2}}{2 \sigma^{2}}\right]\right\}
\label{sexp}
\end{eqnarray}

with mean value zero and ($h_{0}, -h_{0}$), respectively, and
standard deviation $\sigma$ \cite{sneiderpytte,dgaussian}.

Galam and Aharony, in a series of investigations, presented a
detailed analysis via mean field and renormalization group of a
system consisting of $n-$component classical spins (finally choosing $n=3$)
on a $d-$dimensional lattice  of a uniaxially anisotropic
ferromagnet in a longitudinal random field extracted from a
symmetric bimodal PDF ($p=q=1/2$) without and with a uniform
magnetic field along the easy axis, respectively \cite{galamaharony,galam1}. The
uniaxial anisotropy was chosen to be along the easy axis and the
exchange couplings were of the form $J^{(2)}=aJ^{(1)}$, where $a$
is the anisotropy and $0\leq a\leq1$. Depending on the anisotropy
(small, medium, large) a variety of phases (longitudinal,
transverse, paramagnetic), critical, bicritical, critical-end
points as well as a multicritical point (an intersection of
bicritical, tricritical and critical-end-point lines) resulted. In
addition to these purely theoretical investigations, Galam,
proposed a model (diluted random field) in his attempt to
reproduce some of the features in the phase diagram of the
experimental sample consisting of the mixed cyanide crystals
$X(CN)_{x}Y_{1-x}$, where $X$ stands for an alkali metal (K,Na,Rb)
and $Y$ a spherical halogen ion (Br,Cl,I); the dilution of the
pure crystal $XCN$ is achieved by replacing $CN$ by the halogen
ions $Y$ \cite{galam2}. The pure alkali-cyanide $XCN$ crystal
ferroelastic transition disappears at some concentration $x_{c}$
of the cyanide; its numerical value depends on both components
$X,Y$. By choosing a model Hamiltonian (ferromagnetic Ising-type
with nearest neighbor interaction) with dilution and a symmetric
trimodal PDF for the random fields, Galam, using MFA, managed to
predict the involved first and second order phase transitions with
the interfering TCP as well as the respective concentration for a
phase transition to occur depending on the procedure considered.
The random fields were necessary because there were experimental
evidences that below $x_{c}$ cyanide displayed orientational
freezing and the random fields were used for fixing this
orientation. The involved probability $p_{t}$ in PDF as well as
the critical threshold $x_{c}$ were expressed in terms of
microscopic quantities.

Recently, the asymmetric bimodal PDF (\ref{bimodalp}) with $p\neq
q$, in general, has also been studied in detail \cite{asymmetric}.
This study has revealed that for some values of $p$ and $h_{0}$
the PM/FM boundary is exclusively of second order; however, for
some other ranges of these variables this boundary consists of two
branches, a second order one and another of first order, thus
confirming the existence of a tricritical point, whose temperature depends
only on the probability $p$ in (\ref{bimodalp}). In addition to
these findings, the occurrence of reentrance has been
corroborated as well as complex magnetization profiles with
respect the random field strength $h_{0}$. For $p = q$, symmetric
bimodal PDF, the results found by Aharony were confirmed
\cite{aharony}.

An immediate generalization of the asymmetric bimodal
(\ref{bimodalp}) is the asymmetric trimodal one,

\begin{equation}
 P(h_{i})=p\delta(h_{i}-h_{0}) + q \delta (h_{i}+h_{0}) +
 r \delta(h_{i})  \label{trimodal}
\end{equation}

where $p+q+r=1$. In earlier studies, the probabilities $p, q$ had
been considered as equal and related to $r$ by the relation
$p=q=(1-r)/2$, symmetric PDF \cite{trimodal,saxena}. The
third-peak, introduced in addition the other two ones in the
bimodal (\ref{bimodalp}) and associated with the third term in
(\ref{trimodal}), is to allow for the presence of non magnetic
spins or vacancies in the lattice that are not affected by the
random magnetic fields and reduces the randomness of the system,
as well.

For the critical exponents of the three-dimensional RFIM, it seems
that there is broad consensus concerning their values except for
the specific heat exponent $\alpha$, for which there is much
dispute concerning its numerical value, since its sign is widely
accepted to be negative. The main source of information for the
critical exponents are Monte Carlo simulations. However, they
provide various values depending on the probability distribution
considered. Middleton and Fisher concluded that the
$\alpha$-exponent is near zero, $\alpha = -0.01 \pm 0.09$
\cite{middleton}. Rieger and Young, considering the bimodal
distribution, found $\alpha = -1.0 \pm 0.3$ \cite{riegeryoung},
Rieger, using the single Gaussian distribution, found $\alpha =
-0.5 \pm 0.2$ \cite{rieger}, whereas Hartmann and Young, in
ground-state calculations, found $\alpha = -0.63 \pm 0.07$
\cite{hartmannyoung}. Nowak et al concluded that $\alpha = -0.5
\pm 0.2$ \cite{nowak}, whereas Dukovski and Machta found a
positive value, namely, $\alpha = 0.12$ \cite{dukovski}. Malakis
and Fytas \cite{malakisfytas}, by applying the critical
minimum-energy subspace scheme in conjunction with the Wang-Landau
and broad-histogram methods for cubic lattices, proved that the
specific heat and susceptibility are non-self-averaging for $d=3$
using the bimodal distribution. The same ambiguous situation
prevails in experimental measurements, see Ref.
\cite{belangeryoung}.

In this work, we study the RFIM with the asymmetric trimodal PDF
(\ref{trimodal}) with arbitrary values for the probabilities $p,
q$ ($p+q+r=1$) in order to investigate the phase diagrams, phase
transitions, tricritical points and magnetization profiles with
respect to $h_{0}$. The paper is organized as follows: In the next
section, the suitable Hamiltonian is introduced and the respective
free energy and equation of state for the magnetization are
derived. In section $3$, the phase diagram, tricritical points and
magnetization profiles for various values of $p$ and $q$ are
calculated and discussed; we close with the conclusions in section
$4$.

\vspace{-5mm}

\section{The model}

\vspace{-5mm}

The pure Ising model Hamiltonian in the presence of random fields
changes into,

\begin{equation}
 H=-J\sum_{<i,j>}S_{i}S_{j}-\sum_{i}h_{i}S_{i}  \hspace{2mm},
    \hspace{20mm} S_{i}=\pm1 \label{rham}
\end{equation}

The summation in the first term extends over all nearest neighbors
and is denoted by $<i,j>$; in the second term $h_{i}$ represents
the random field that couples to the one-dimensional spin variable
$S_{i}$. We also consider that $J > 0$ so that the ground state is
ferromagnetic in the absence of random fields. The presence of
randomness necessitates considering two averaging procedures, the
usual thermal average, denoted by angular brackets
$\langle...\rangle$, and the disorder average over the random
fields denoted by $\langle...\rangle_{h}$ for the respective PDF.
We also make assumptions concerning the random field $h_{i}$,

\begin{equation}
 <h_{i}>_{h} = (p-q)h_{0},
   \hspace{20mm} <h_{i} h_{j}>_{h} = h^{2}_{0}\delta _{ij} \label{h0}
\end{equation}

The former relation in (\ref{h0}) vanishes for a symmetric PDF
($p=q$), whereas for the asymmetric PDF ($p \neq q$) it is non
zero implying that the system has residual magnetization, thereby
affecting considerably the system's magnetization; a similar
case has appeared in Ref. \cite{asymmetric}. The latter one implies
that there is no correlation between $h_{i}$ at different lattice sites.

According to the MFA, the Hamiltonian (\ref{rham}) takes the form
\cite{aharony,sneiderpytte,andelman1,asymmetric},

\begin{equation}
H_{MFA}=\frac{1}{2} NzJM^{2} - \sum_{i}(zJM + h_{i})S_{i}
     \label{mfaham}
\end{equation}

where $N$ is the number of spins in the lattice and $M$ the
magnetization; the respective free energy per spin within the MFA is,

\begin{eqnarray}
\frac{1}{N}\langle F \rangle_{h} & = & \frac{1}{2} zJM^{2} -
\frac{1}{\beta} \langle \ln\{ 2 \cosh [\beta (z J M + h_{i})] \}
\rangle _{h}          \nonumber  \\
  & = &  \frac{1}{2} zJM^{2} - \frac{1}{\beta}
 \int P(h_{i})\ln\{ 2 \cosh [\beta (z J M + h_{i})] \} dh_{i}
                \label{mfafren}
\end{eqnarray}

where the probability $P(h_{i})$ is chosen to be the trimodal
(\ref{trimodal}) and $p, q$ take on any value within the interval
[0,1], consistent with the relation $p+q+r=1$, and $\beta=1/(kT)$.

The magnetization is the solution to the equation $d(\langle F
\rangle_{h}/N) / dM = 0$, equilibrium condition,

\begin{equation}
  M = \langle \tanh [ \beta ( zJM + h_{i} ) ] \rangle_{h}
     \label{magnet1}
\end{equation}

If the distribution $P(h_{i})$ is symmetric, $P(h_{i})$ =
$P(-h_{i})$, which occurs for $p = q = (1-r)/2$, then the value $M
= 0$ (PM phase) will always be a solution to (\ref{magnet1}),
otherwise this is not if $P(h_{i})$ is non symmetric, $p \neq q$;
however, this can be remedied if an auxiliary field $V_{0}$ is
introduced into the system such that \cite{aharony,asymmetric},

\begin{equation}
  \langle \tanh[ \beta ( h_{i} + V_{0} ) ] \rangle_{h} = 0
  \label{externalv}
\end{equation}

inducing, in this way, the PM phase. However, this relation acts
also as a constraint on the system under consideration influencing,
nevertheless, its behavior. The free energy (\ref{mfafren}), in
the presence of the auxiliary field $V_{0}$, takes now the form,

\begin{eqnarray}
\frac{1}{N}\langle F \rangle_{h} & = & \frac{1}{2} zJM^{2} -
\frac{1}{\beta} \langle \ln\{ 2 \cosh [\beta (zJM + h_{i} +
V_{0})] \}
\rangle_{h}          \nonumber  \\
  & = &  \frac{1}{2} zJM^{2} - \frac{1}{\beta}
 \mbox{\Large \{}\! F_{0} + \frac{\alpha ^{2} F_{2}}{2!} M^{2} +
 \frac{\alpha ^{3} F_{3}}{3!}M^{3}
 +\frac{\alpha ^{4} F_{4}}{4!} M^{4} + \nonumber  \\
&& \frac{\alpha ^{6} F_{6}}{6!}M^{6}  \mbox{\Large \}}
\label{mfafren2}
\end{eqnarray}

after expanding the quantity in angular brackets in powers of $M$
and calculating the average values using (\ref{trimodal}) with $\alpha
\equiv \beta Jz$. By setting $t_{i} \equiv \tanh[\beta(V_{0}+h_{i})]$,
$t_{+} \equiv \tanh[\beta(V_{0}+h_{0})]$, $t_{-} \equiv
\tanh[\beta(V_{0}-h_{0})]$ and $t_{0} \equiv \tanh[\beta V_{0}]$, we have,

\begin{eqnarray}
F_{0} & = & \langle \ln\{ 2\cosh[ \beta ( V_{0} + h_{i} )  ] \}
\rangle_{h}     \nonumber  \\
  & = & \ln2 + p\ln\cosh[\beta (V_{0} + h_{0})] + q\ln\cosh[\beta (V_{0} - h_{0})]  \nonumber
      + r\ln\cosh[\beta V_{0} ]  \nonumber  \\
F_{1} & = & \langle t_{i}\rangle_{h} =  p t_{+} + q t_{-} + r t_{0}   \nonumber  \\
F_{2} & = & \langle 1 - t_{i}^{2}\rangle_{h} =  1 - p t_{+}^{2} - q t_{-}^{2}-r t_{0}^{2}  \nonumber \\
F_{3} & = & \langle -2t_{i} (1-t_{i}^{2}) \rangle_{h}   \nonumber \\
  & = & -2p t_{+} (1-t_{+}^{2}) -2 q  t_{-} (1-t_{-}^{2})-2 r t_{0}(1-t_{0}^{2})  \nonumber\\
F_{4} & = & \langle 2 (1 - t_{i}^{2}) (3t_{i}^{2}-1) \rangle_{h}   \nonumber \\
  & = & 2p (1 - t_{+}^{2}) (3t_{+}^{2}-1) + 2q (1 - t_{-}^{2}) (3t_{-}^{2}-1)   \nonumber
      + 2r (1 - t_{0}^{2}) (3t_{0}^{2}-1)   \nonumber \\
F_{6} & = & \langle 8 (1 - t_{i}^{2})(15t_{i}^{4}-15t_{i}^{2}+2)
\rangle_{h}  \nonumber \\
  & = & 8p (1 - t_{+}^{2})(15t_{+}^{4}-15t_{+}^{2}+2)+
        8q (1 - t_{-}^{2})(15t_{-}^{4}-15t_{-}^{2}+2)  \\  \nonumber
  &  &  +8r (1 - t_{0}^{2})(15t_{0}^{4}-15t_{0}^{2}+2)   \label{mfafren3}
\end{eqnarray}

The condition (\ref{externalv}) for the existence of the PM phase
is equivalent to setting $F_{1} = 0$,

\begin{equation}
  p t_{+} + q t_{-} + r t_{0} = 0   \label{f1zero}
\end{equation}

The equilibrium condition $d(\langle F \rangle_{h}/N) / dM = 0$
yields,

\begin{equation}
M=\alpha F_{2} M + \frac{\alpha ^{2} F_{3}}{2!} M^{2} +
   \frac{\alpha ^{3} F_{4}}{3!} M^{3} +
   \frac{\alpha ^{5} F_{6}}{5!}M^{5}            \label{eqmagn}
\end{equation}

\vspace{-3mm}

or,

\vspace{-3mm}

\begin{eqnarray}
M & = & A M + B M^{2}+ C M^{3} + E M^{5}  \label{magnet1a} \\
A & \equiv & \alpha F_{2}, B\equiv\frac{\alpha ^{2} F_{3}}{2!},
C\equiv\frac{\alpha ^{3} F_{4}}{3!}, E\equiv\frac{\alpha ^{5}
F_{6}}{5!}         \label{magnet2}
\end{eqnarray}

In RFIM if there is a phase transition it will be associated with
the magnetization and the possible two phases are the PM with
$M=0$ and FM with $M\neq0$. The phase boundary is found by solving
Eq. (\ref{magnet1a}) in conjunction with the free energy
(\ref{mfafren2}) in case of an FOPT. The SOPT boundary is
determined by setting $A = 1$ and $C < 0$, whereas the FOPT
boundary by $A = 1$ and $C > 0$. These two boundaries, whenever
appear sequentially, are joined at a tricritical point determined
by the condition $A = 1$ and $C= 0$
\cite{aharony,andelman1,kaufkan,dgaussian,asymmetric,crok2,crok3,khurana},
provided that $E<0$ (equivalently, $F_{6}<0$) for stability, as also in
\cite{crok2,crok3}. However, for the FOPT boundary we shall use,
in addition to (\ref{f1zero}) and (\ref{magnet1a}), the requirement
of the equality of the respective free energies, $F(M=0) = F(M
\neq 0)$, where $F \equiv \langle F \rangle_{h}/N$.

\vspace{-5mm}

\section{Phase diagram. Magnetization profiles}

\vspace{-5mm}

\begin{figure}[htbp]
\includegraphics*[height=0.20\textheight]{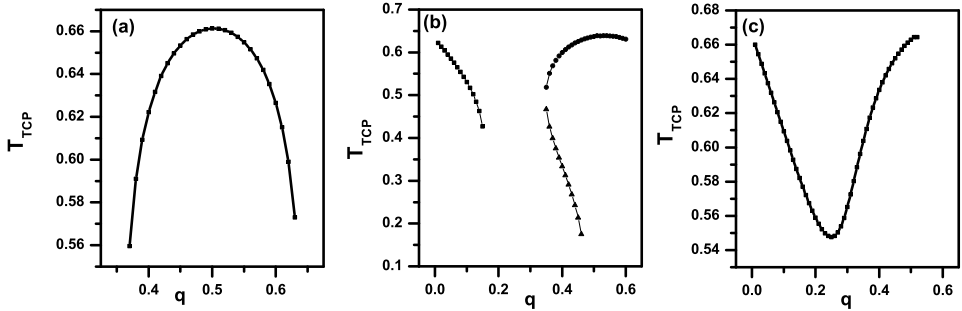}
\caption{\label{figa} The tricritical temperature against the
probability $q$ for specific values of $p$. In panel (a) ($p =
0.01$) it varies non-monotonically with a maximum value at
$q=0.50$. In panel (b) ($p=0.40$) the left-hand-side points
correspond to tricritical temperatures for low q-values ($0.0 \leq
q \leq 0.15$). The right-hand-side group of points forms two
branches, the upper one refers to the upper tricritical
temperatures ($0.35 \leq q \leq 0.60$), whereas the lower one
corresponds to the lower tricritical temperatures for fewer values
of $q$, $0.35 \leq q \leq 0.46$. In panel (c) ($p=0.48$) the
tricritical temperature varies non-monotonically with a minimum at
$q=0.25$.}
\end{figure}

Using the conditions for the calculation of the TCP together with
Eq. (\ref{f1zero}), the TCP coordinates $(T^{TCP}, h^{TCP}_{0},
V^{TCP}_{0})$ are calculated as functions of the probabilities $p,
q$. These exist only for a limited number of $p$'s and $q$'s,
namely, $p \in [0.0,0.63]$ whereas the respective $q-$values
depend on the specific $p-$values, but they lie in same interval,
as well. However, the tricritical temperature $T^{TCP}$ does not
satisfy any simple closed-formula as in the asymmetric bimodal PDF
\cite{asymmetric}. The resulting tricritical temperatures exhibit
a variety of variations as functions of $p, q$, see
Fig.~\ref{figa}. However, for some p- and q-values two tricritical
temperatures (upper and lower ones) occur, see Fig.~\ref{figa}(b)
\cite{galam,weizenmann}, since both temperatures are solutions to
the simultaneous equations $ \alpha F_{2}=1, F_{4}=0$.

\begin{figure}[htbp]
\includegraphics*[height=0.35\textheight]{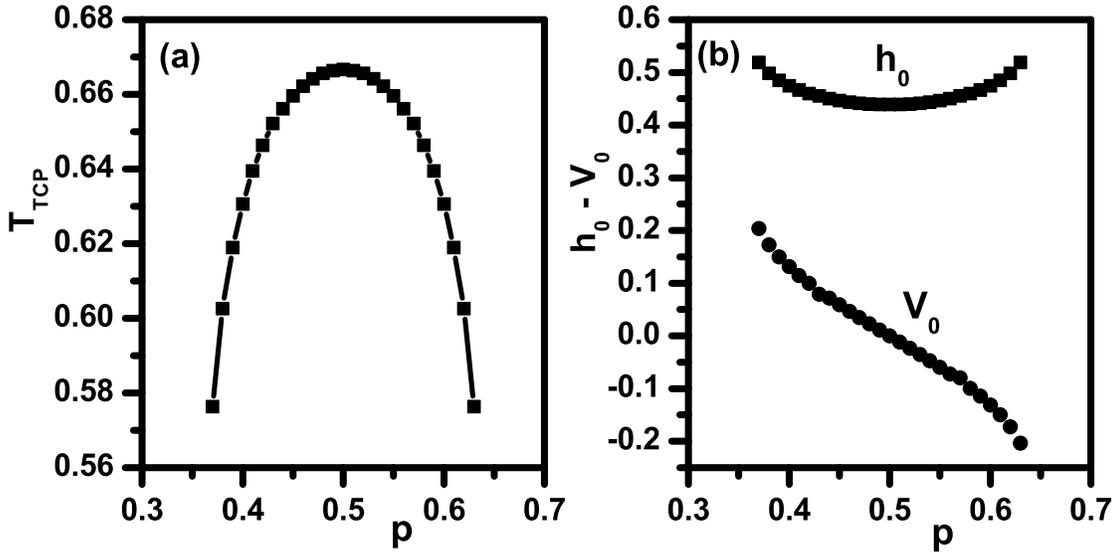}
\caption{\label{figb} Tricritical point coordinates for the
bimodal PDF, (a) tricritical temperature and (b) random field
$h_{0}$ and auxiliary field $V_{0}$, resulting form trimodal PDF
in the limiting case $r=0$. The agreement of these figures with
the corresponding ones in Ref.~\cite{asymmetric} is complete.}
\end{figure}

\begin{figure}[htbp]
\includegraphics*[height=0.35\textheight]{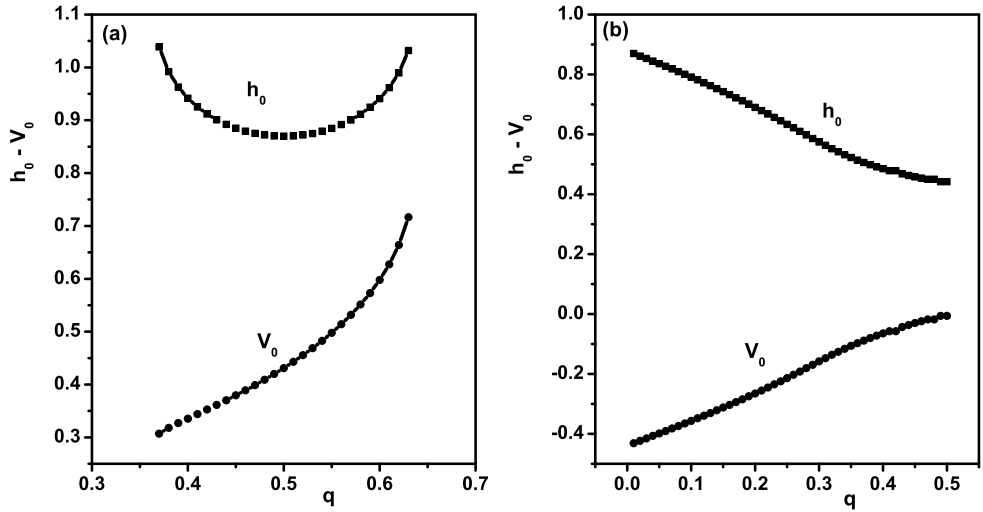}
\caption{\label{figc} Variation of the random field strength
$h_{0}$ and auxiliary field $V_{0}$ with $q$ for specific
$p-$values at the tricritical point for the trimodal PDF. The
random field $h_{0}$ exhibits a non-monotonic behavior in (a) for
$p=0.01$ and monotonic in (b) for $p=0.50$. The auxiliary
potential $V_{0}$ increases monotonically with $q$ for any $p$.
Both quantities are in units of $(Jz)$, i.e., $h_{0} \equiv
h_{0}/(Jz)$, $V_{0} \equiv V_{0}/(Jz)$.}
\end{figure}

In order to examine the validity of the process under
consideration, we focus on the asymmetric bimodal PDF resulting
from the trimodal by setting $r=0$ and studied earlier
\cite{asymmetric}; using the data for $r=0$, we recover exactly
the same plots for the tricritical temperature as well as $h_{0}
,V_{0}$, see Fig.~\ref{figb}.

Now, using the relations $A=1$ and $C=0$ (equivalently, $\alpha
F_{2}=1$ and $F_{4}=0$) together with (\ref{f1zero}), the
remaining coordinates $h_{0}^{TCP}$ and $V_{0}^{TCP}$ for the TCP
are calculated, see Fig.~\ref{figc}. Both quantities display two
modes of variation; in mode-$1$ ($p \leq 0.49$) $h_{0}^{TCP}$
varies non-monotonically with $q$, Fig.~\ref{figc}(a); in mode-$2$
($p \geq 0.50$) $h_{0}^{TCP}$ decreases monotonically with $q$,
Fig.~\ref{figc}(b). However, in both modes, $V_{0}^{TCP}$
increases monotonically, but for type-$1$ the increase is more
steep than in type-$2$.

The magnetization at the TCP is found by solving Eq.
(\ref{eqmagn}) taking into consideration the appropriate
conditions for the TCP,

\begin{equation}
\frac{\alpha ^{2}F_{3}}{2!}M^{2} + \frac{\alpha ^{5}
F_{6}}{5!}M^{5} = 0 \label{tcpmagn1}
\end{equation}

or

\begin{equation}
F_{6} \omega^{5} + 60 F_{3} \omega^{2} = 0 \label{tcpmagn2}
\end{equation}

\vspace{-5mm}

where $\omega \equiv \alpha M$, whose solutions are,

\begin{eqnarray}
\omega _{1}^{TCP} & = & 0  \label{tcpmagn0}    \\
\omega _{2}^{TCP} & = & \mbox{\Large(}-60 F_{3} /F_{6}
\mbox{\Large)}^{1/3}       \label{tcpmagn3}
\end{eqnarray}

\begin{figure}[htbp]
\includegraphics*[height=0.4\textheight]{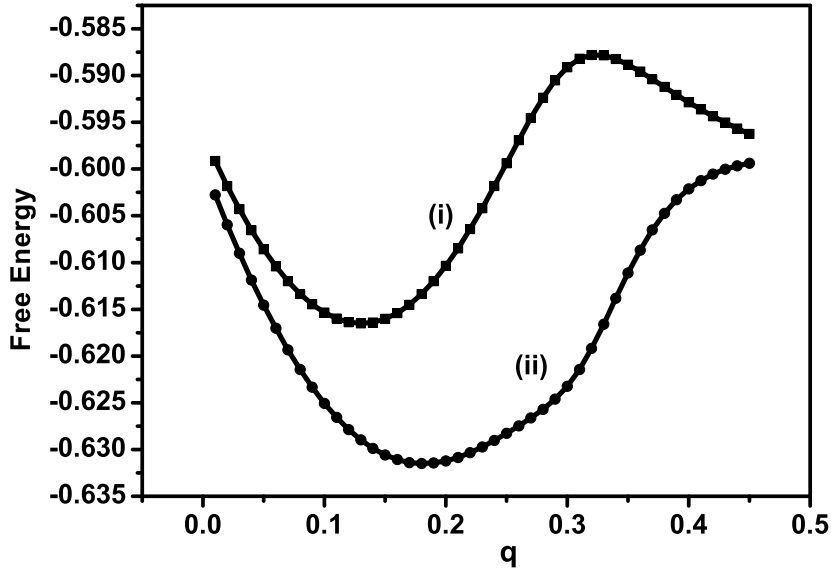}
\caption{\label{figd} Free energy of the solutions
(\ref{tcpmagn0}) and (\ref{tcpmagn3}) at the tricritical point for
$p=0.55$; the upper plot (i) corresponds to the zero solution
$M_{1}$ and the lower (ii) to the nonzero solution $M_{2}$. For
this $p-$value, the $M_{2}$ is the stable solution, whereas the
$M_{1}$ is metastable; this happens, in general, for other
$p-$values. The $M_{2}$ solution coincides with the zero one
($M_{1}$) only for $p=q$.}
\end{figure}

\begin{figure}[htbp]
\includegraphics*[height=0.40\textheight]{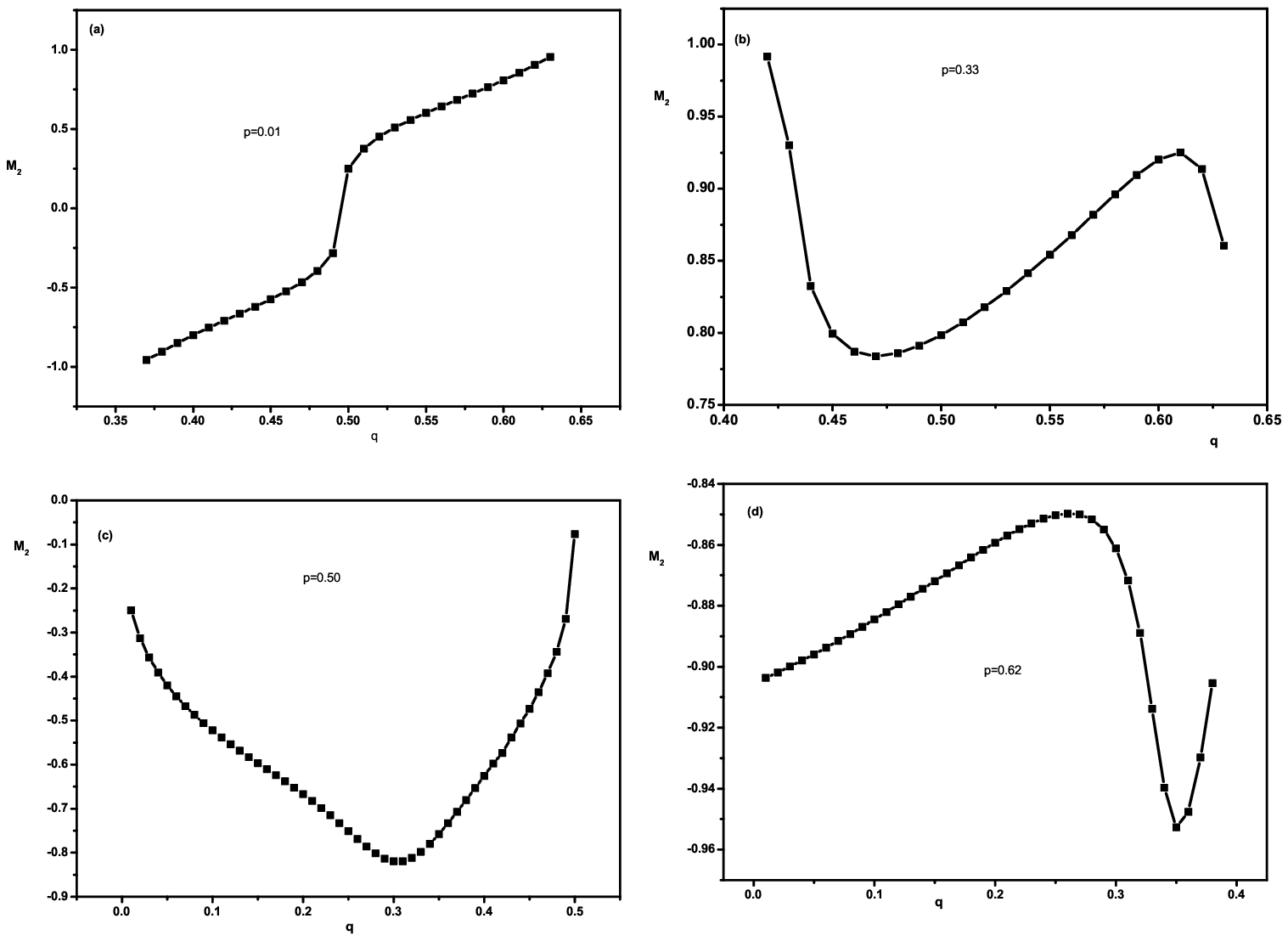}
\caption{\label{fige} Variation of the nonzero magnetization
$M_{2}^{TCP}= \mbox{\large(}-60 F_{3} /F_{6} \mbox{\large)}^{1/3}$
with $q$ for specific $p-$values at the tricritical point. $M_{2}$
is, in general, the stable solution. Except for $p=0.01$, in the
other panels the TCP-magnetization displays a non monotonic
behavior as a function of $q$.}
\end{figure}

from which the magnetizations $M_{1,2}^{TCP} = \omega
_{1,2}^{TCP}*(kT^{TCP}/(Jz))$ can be deduced. The first one,
(\ref{tcpmagn0}), is the magnetization of the PM phase
($M_{1}^{TCP} = 0$), whereas the second one, (\ref{tcpmagn3}), is
the magnetization of the FM phase ($M_{2}^{TCP} \neq 0$). However,
the nonzero solution $M_{2}^{TCP}$, in general, has lower free
energy than the zero solution (\ref{tcpmagn0}), implying that this
is the stable solution at the tricritical point, see Fig.~\ref{figd}.
According to the first relation in Eqs. (\ref{h0}), for the general
case $p \neq q$, the mean value of the random field is non zero;
this is equivalent to the presence of an external magnetic field in
the system so that the magnetization at the tricritical point scales
as $M_{t}\equiv M(T=T^{TCP}) \sim h^{1/\delta_{t}}_{TCP}$, where $h_{TCP}$
is the magnetic field and the tricritical exponent $\delta _{t} = 5$
according to the Landau theory \cite{stanley,robertson,lawrie}.
In case of equal partial probabilities ($p=q$), the
$M_{2}^{TCP}$-magnetization vanishes ($M_{2}^{TCP}=0=M_{1}^{TCP}$)
and the system has only the zero solution (double root), which now
becomes the stable one. Representative plots of the magnetization
$M_{2}^{TCP}$ with respect to $q$ appear in Fig.~\ref{fige} for
specific values of $p$.

In a previous communication \cite{asymmetric}, the PDF of the RFIM
was selected to be the asymmetric bimodal (\ref{bimodalp}); this
system displayed a symmetric behavior at the tricritical point
with respect to the probability $p$\,; especially, two distinct
tricritical points with probabilities $p_{1}$ and $p_{2}$,
respectively, such that $p_{1}+ p_{2} = 1$, have identical
tricritical temperatures and random fields, whereas the respective
auxiliary fields and non zero magnetizations are absolutely equal.
A similar behavior is also displayed by the present model with
respect to the probabilities $p$ and $q$; if the probabilities
($p_{1},q_{1}$) and ($p_{2},q_{2}$) of the tricritical points of
two systems are interchanged, namely, $p_{2}=q_{1}, q_{2}= p_{1}$,
then these systems have the same tricritical temperatures and
random fields, whereas the respective auxiliary fields
$V_{0}^{TCP}$ and the nonzero magnetizations $M_{2}^{TCP}$ are
absolutely equal. These systems also have equal the respective
free energies for the zero magnetization
($F(p_{1},q_{1},M_{1}^{TCP}=0) = F(p_{2},q_{2},M_{1}^{TCP}=0)$) as
well as the nonzero one ($F(p_{1},q_{1},M_{2}^{TCP}) =
F(p_{2},q_{2},M_{2}^{TCP})$); the latter result implies that the
two magnetizations $M_{2}^{TCP}(p_{1},q_{1})$,
$M_{2}^{TCP}(p_{2},q_{2})$ are equally probable, an expected
result, since the magnetizations have equal absolute values, the
only difference being in their sign, implying that no direction is
favored.

\begin{figure}[htbp]
\includegraphics*[height=0.30\textheight]{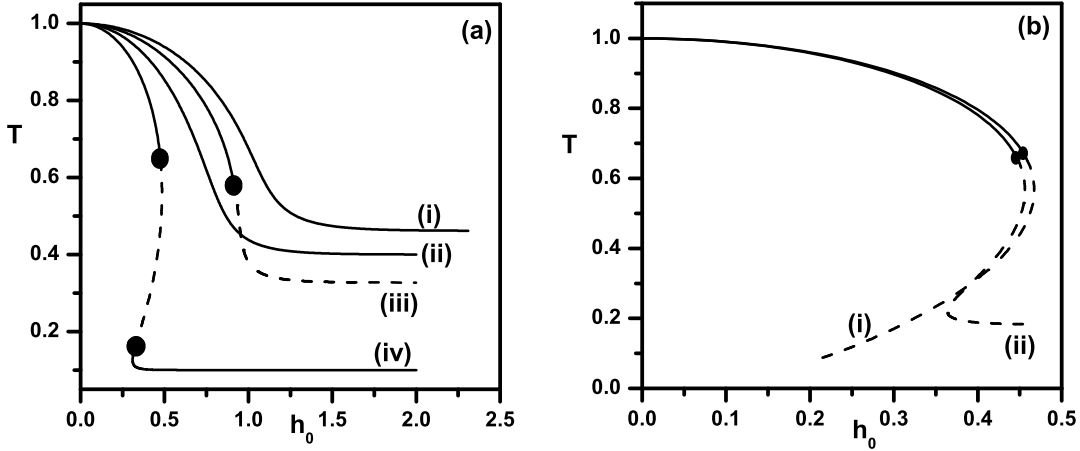}
\caption{\label{figf} Phase diagram of the Hamiltonian
(\ref{rham}) in the ($h_{0}-T$) plane for specific values of $p$
and $q$. The solid curve is a line of critical points and the
dashed one is a line of first-order phase transitions, joined
smoothly by a tricritical point (full circle). The system has, in
general, second-order transition as in panel (a) for
$p=0.35,q=0.0$, curve (i) and $p=0.35, q=0.20$, curve (ii);
however, for $p=0.05, q=0.40$, curve (iii), it displays both phase
transitions, FOPT and SOPT joined by a tricritical point, whereas
for $p=0.45, q=0.45$, curve (iv), the system has two tricritical
points, occurring also for other values of $p$ and $q$. In the
latter case, the system displays reentrance, as well. In panel (b)
the phenomenon of reentrance is shown in enlargement; for $p=0.45,
q=0.50$ (curve (i)) the system remains in the PM phase for low
temperatures and medium random fields, whereas for $p=0.45,
q=0.55$ (curve (ii)) it returns to the FM phase for low
temperatures and higher random fields. Temperature $T$ is
expressed in units of $(Jz/k)$, i.e., $T \equiv kT/(Jz)$.}
\end{figure}

After the calculation of the tricritical point coordinates
($T^{TCP}, h_{0}^{TCP}, V_{0}^{TCP}$) for the respective values of
$p$ and $q$, we proceed to determine the phase diagram by solving
Eq. (\ref{magnet1a}), taking into consideration the respective
conditions for the FOPT and the SOPT. By varying the parameters
$p$ and $q$ many different types of phase diagrams result, as
those appearing in Fig.~\ref{figf}(a) labelled by the individual
$p$ and $q$ values. The curves in Fig.~\ref{figf}(a) are
classified into two main groups: the first one includes those
curves that do not possess a TCP corresponding to an SOPT only,
curves (i) and (ii); the second group includes those having a TCP,
which joins the FOPT-branch with the SOPT-branch of the phase
diagram, curves (iii) and (iv). The occurrence of an FOPT, and
subsequently of a TCP, results from the competition between the
first term in the Hamiltonian (\ref{rham}) (tending to make
parallel the spins) and the second term of random forces inducing
disorder. For small values of $h_{0}$ the competition is weak
allowing the first term to dominate, but as $h_{0}$ increases the
second term dominates over the first one, thus changing the phase
transition from second order to first order. Some of the curves in
the latter group have a second tricritical point and/or present
reentrance Fig.~\ref{figf}(a(iv)). Reentrance might also be
attributed to the competition between these two terms in the
Hamiltonian (\ref{rham}). In the phenomenon of reentrance a
vertical line in the $(h_{0},T)$-plane crosses the transition line
twice, in that, by lowering the temperature at constant $h_{0}$,
one observes first a $PM/FM$ transition and then, on further
lowering the temperature, an $FM/PM$ transition appears so that
the magnetization is zero although the temperature is low and the
system remains in the PM phase for these temperatures,
Fig.~\ref{figf}(b(i)) or another transition may take place from
$PM$ to $FM$ with the system, now, in the $FM$ phase for low
temperatures and high fields, Fig.~\ref{figf}(b(ii));
occasionally, the region of the FM phase shrinks significantly,
Fig.~\ref{figf}(a(iv)). The vanishing of magnetization for high
values of $p$ ($p\sim 0.45$, $45\%$ up-spins) and small $q$ values
(where one would expect a nonzero magnetization) can also be due
to the presence of the auxiliary field $V_{0}$, which annihilates
the excess magnetization making the system to behave like an
antiferromagnet. This phenomenon is evident from the bending of
the phase transition lines on lowering the temperature thus
forming an inverted "C" ("boomerang" shape) and appearing in
enlargement in Fig.~\ref{figf}(b). The respective p values lie in
the interval $[0,0.50]$, whereas the q values lie in a much
smaller interval, namely, $[0.40,0.55]$; the only exception is for
$p=0.45$ when $q$ takes on values in the interval $[0,0.55]$.
However, within the MFA reentrance may lead to nonphysical values
(negative) for the specific heat, since energy will also present
reentrant behavior as magnetization because energy, within MFA, is
proportional to the magnetization squared thus behaving similarly.
The two successive transitions can be of first or second order
depending on $p, q$ and  $h_{0}$.

Solving Eq. (\ref{magnet1a}) to determine the phase diagram, the
magnetization is also calculated for either phase transition. The
condition $A = 1$ or $\alpha F_{2}(\beta,V_{0},h_{0}) = 1$ leads
to

\vspace{-3mm}

\begin{equation}
p\,t_{+}^{2} + qt_{-}^{2}+ rt_{0}^{2} = \frac{\alpha -1}{\alpha}
\;\;\;\; \;\; \label{t2a}
\end{equation}

\vspace{-3mm}

and, by setting $T_{2} \equiv  p\,t_{+}^{2} + qt_{-}^{2}+
rt_{0}^{2}$, Eq. (\ref{t2a}) can be written as,

\vspace{-3mm}

\begin{equation}
  T_{2} = \frac{\alpha -1}{\alpha} \;\;\;\; \;\; \label{t2b}
\end{equation}

\vspace{-3mm}

Inverting Eq. (\ref{t2b}) the respective temperature for either
phase transition can be determined,

\vspace{-3mm}

\begin{equation}
  \frac{kT}{Jz} = 1 - T_{2}   \;\;\;\; \;\; \label{temp}
\end{equation}

\vspace{-5mm}

In order to specify the type of the transition, the sign of $C
\equiv \alpha^{3} F_{4}/6$ is checked; however, to facilitate the
calculations, the quantity $C$ is rewritten as,

\vspace{-4mm}

\begin{equation}
C = \frac{\alpha^{3}}{3}[4T_{2} - 3T_{4} - 1]
=\alpha^{3}[1-T_{4}-\frac{4}{3\alpha}]  \;\; \;\; \label{consc}
\end{equation}

\vspace{-5mm}

using Eq. (\ref{t2b}) and $T_{4} = p\,t^{4}_{+} + q\,t^{4}_{-}+
r\,t^{4}_{0}$. For an SOPT, $C$ is negative
\cite{aharony,sneiderpytte,asymmetric}, then Eq. (\ref{consc})
yields,

\begin{equation}
T_{4} > 1 - \frac{4}{3\alpha}  \;\;\;\; \;\; \label{sectrns}
\end{equation}

\vspace{-5mm}

otherwise if

\vspace{-5mm}

\begin{equation}
T_{4} < 1 - \frac{4}{3\alpha}  \;\;\;\; \;\; \label{firsttrns}
\end{equation}

\vspace{-5mm}

the resulting transition is an FOPT. However, in order to
determine the magnetization for an FOPT the expression
(\ref{eqmagn}) is combined with the equality of the respective
free energies,

\begin{equation}
 F(M=0)  =  F(M\ne 0) \label{feeq1}
\end{equation}

\vspace{-5mm}

or,

\vspace{-5mm}

\begin{equation}
M^{2}=F_{2} \alpha M^{2}+\frac{F_{3}}{3}\alpha^{2}M^{3}+
\frac{F_{4}}{12} \alpha^{3} M^{4}+\frac{F_{6}}{360} \alpha^{5} M^{6}
\label{feeq2}
\end{equation}

\vspace{-5mm}

Combining Eqs. (\ref{eqmagn}), (\ref{feeq2}) and using the
condition $\alpha F_{2} =1$, leads to,

\begin{equation}
F_{6}\omega^{3} + 10F_{4}\omega = 0 \label{fopt2}
\end{equation}

which is split up into two equations, the first is $\omega _{1} =
0$ or, equivalently, $M_{1} = 0$, PM phase, whereas the other is,

\begin{equation}
F_{6}\omega^{2} + 10 F_{4}= 0 \label{fopt3}
\end{equation}

\vspace{-5mm}

leading to the nonzero solutions, FM phase,

\begin{equation}
 \omega_{2,3}= \pm \sqrt{-10 F_{4}/F_{6}}   \label{foptsols}
\end{equation}

for which $F_{6} < 0$ for stability requirements as far as $F_{4}
> 0$ for an FOPT; the value for $F_{3}$, consistent with
(\ref{eqmagn}) and (\ref{feeq2}), is $F_{3} = - (F_{4}/6)
\sqrt{-10 F_{4}/F_{6}}$ for the positive root in (\ref{foptsols})
and $F_{3} = (F_{4}/6) \sqrt{-10 F_{4}/F_{6}}$ for the respective
negative root. From the solution of this equation we can extract
the magnetizations $M_{2,3}$, since the temperature $(kT/zJ)$ is
already known from (\ref{temp}).

\begin{figure}[htbp]
\includegraphics*[height=0.35\textheight]{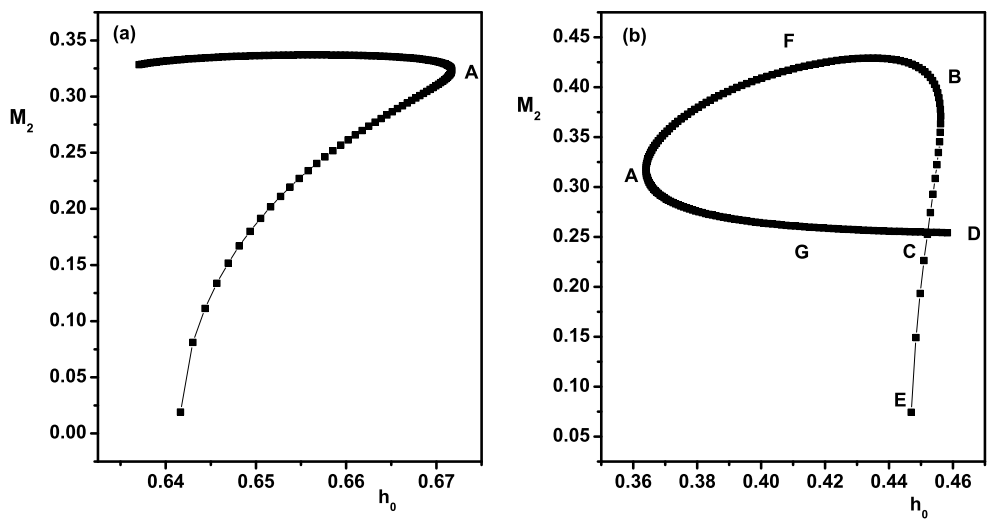}
\caption{\label{figg} Magnetization profile vs. $h_{0}$, for (a)
$p=0.25, q=0.55$, point A is a critical point. (b) $p=0.45,
q=0.55$ with three critical points A, B, C. Point A is a normal
critical point. B a critical-end-point; the two critical phases
are on either branch at point B coexisting with the non critical
phase on the branch CD. C a double critical-end-point; the first
group of the two critical phases are on the branches CGA and CE
coexisting with the non critical phase on the branch BF, the
second group of two critical phases are on the branches CD and CB
coexisting with the non critical phase on the branch BF. This plot
represents a closed-loop phase diagram with left-hand and
right-hand-side critical points.}
\end{figure}

Considering, now, the randomness strength $h_{0}$ as a control
parameter, similarly as temperature, we study the variation of the
non-zero positive magnetization $M_{2} = \omega_{2}*(kT/(Jz))$ for
an FOPT, Eq. (\ref{foptsols}), coexisting in equilibrium with the
zero magnetization $M_{1} = 0$, by calculating the respective
magnetization profile as a function of $h_{0}$ for specific values
of $p$ and $q$; the negative $M_{3} = - M_{2}$ behaves
analogously. These profiles appear in Fig.~\ref{figg} as functions
of $h_{0}$ and look like the ones when the temperature is the
control parameter for an FOPT. For $p=0.25, q=0.55$ the
magnetization has a normal critical point Fig.~\ref{figg}(a),
whereas for $p=0.45, q=0.55$ this structure is transformed into a
closed loop magnetization profile, closed miscibility gap,
Fig.~\ref{figg}(b); it possesses multiple critical points, the
point $A$ is a pure critical point, whereas $B$ is a
critical-end-point (CEP) because the thermodynamic states along the
branches BF and BC become identical at the point B in the presence
of the third non critical phase (spectator phase) along the branch
CD for the same value of $h_{0}$; the CEPs can be found in binary
fluid mixtures, superfluids, binary alloys, liquid crystals,
ferromagnets, ferroelectrics, etc.
\cite{asymmetric,hadjievans,plalan,tsai,rowwid,rowswi,widom,kumar,kraska}.
A simple example for the occurrence of the CEP is to consider a
binary fluid mixture with the proper thermodynamic conditions so
that three fluid phases can coexist in equilibrium and enclosed in
a capsule; these phases are called $V$ (vapor, top phase), $L_{1}$
(liquid phase rich in species 1, middle phase) and $L_{2}$ (liquid
phase rich in species 2, bottom phase). The top and middle phases
are separated by the interface ($VL_{1}$), as well as the middle
and the bottom ones by the interface ($L_{1}L_{2}$); if the
thermodynamic conditions are such that the interface $VL_{1}$
disappears with the phases $V$, $L_{1}$ becoming identical in the
presence of the non critical $L_{2}$-phase, then that
thermodynamic state is a CEP at the temperature $T^{CEP}_{VL_{1}}$
and the new phase $VL_{1}$ coexists with the $L_{2}$ one in
equilibrium. However, if we consider that the interface
$L_{1}L_{2}$ disappears with the phases $L_{1}$, $L_{2}$ becoming
identical in the presence of the non critical $V$-phase, then that
thermodynamic state is a CEP at the temperature
$T^{CEP}_{L_{1}L_{2}}$ and the new phase $L_{1}L_{2}$ coexists
with the $V$ one in equilibrium. Another occurrence of a CEP is
the double critical end point (DCEP) where two critical lines end
simultaneously at a first order phase boundary; such a situation
appears at the point $C$, where  the two critical phases on the
branches CGA and CE become identical at the point C in the
presence of the non critical phase on the branch BF; the second
group of two critical phases are on the branches CD and CB that
become identical in the presence of the non critical phase on the
branch BF; these are also observed in binary fluid mixtures,
metamagnets, Ising antiferromagnets with next-nearest-neighbor
interactions, RFIM, three dimensional antiferromagnetic spin-1
Blume-Capel model \cite{asymmetric,plalan,kraska,wangkimel}. Point
$A$ can also be considered as a left-hand side critical point and
$B, C$ right-hand side ones.

Now, we consider the values of $p$ and $q$ for which the system
exhibits only an SOPT; Eq. (\ref{magnet1a}) takes the form for
$A=1$,

\begin{equation}
F_{6}\omega^{5} + 20F_{4}\omega^{3} + 60F_{3}\omega^{2} = 0
\label{sopt1}
\end{equation}

The value $\omega_{1} = 0$ (two-fold) is again a solution or,
equivalently, $M_{1} = 0$ (PM phase); the other three roots are
the solutions to the third degree equation,

\begin{equation}
F_{6}\omega^{3} + 20F_{4}\omega + 60F_{3} = 0 \label{sopt2}
\end{equation}

\vspace{-5mm}

which, depending on the values of $p, q$ and $h_{0}$, can have
either only one real non zero solution if $\Delta = q^{3} + r^{2}
\geq 0$ ($r = -30F_{3}/F_{6}$, $q=(20F_{4})/(3F_{6})$),

\begin{equation}
\omega_{2}=\sqrt[3]{r+\sqrt{\Delta}} + \sqrt[3]{r-\sqrt{\Delta}}
\label{root2}
\end{equation}

or three real non zero ones for $\Delta < 0$,

\vspace{-5mm}

\begin{eqnarray}
\omega_{2} & =& 2 \sqrt[3]{\rho}\;\cos(\theta/3)    \nonumber   \\
\omega_{3} & = & -\sqrt[3]{\rho}\;[\cos(\theta/3) +
\sqrt{3}\sin(\theta/3)]                            \nonumber \\
\omega_{4} & = & -\sqrt[3]{\rho}\;[\cos(\theta/3) -
\sqrt{3}\sin(\theta/3)]                            \label{root34}
\end{eqnarray}

where $\rho = \sqrt{r^{2} - \Delta}$, $\theta = \arctan(
\sqrt{-\Delta}/r )$ and $M_{i} = \omega_{i}*(kT/(Jz))$, $i=2,3,4$.
The solutions for an SOPT are classified into two groups, group-1
includes the zero solution ($M_{1}=0$) and the single nonzero one
$M_{2}$ of (\ref{root2}), whereas group-2 includes again the zero
solution and the nonzero ones $M_{2}, M_{3}, M_{4}$ of
(\ref{root34}). Depending on the value of $p, q$ and $h_{0}$,
there can be transitions between these two groups. The majority of
the SOPT solutions belong to the group-2; for given values of $p$
and $q$ the stable solution is the zero (PM phase) for small
values of $h_{0}$, whereas for higher ones the most probable to be
stable is the $M_{2}$ solution. For the group-1, the zero solution
is stable for small values of $h_{0}$ and the $M_{2}$ solution for
higher values.

The investigation was also extended towards the zero-temperature
case, $T=0$; in this case the free energy (\ref{mfafren}) reduces
to,

\begin{eqnarray}
 \frac{1}{N}\langle F \rangle_{h} & = & \frac{1}{2}
zJM^{2} - \frac{1}{\beta} \langle \ln\{ 2 \cosh [\beta (z J M +
h_{i})] \}
\rangle_{h}          \nonumber  \\
  &  =  &  \frac{1}{2} zJM^{2} - \langle |z J M + h| \rangle_{h}    \nonumber  \\
  & = & \frac{1}{2} zJM^{2} - p|z J M + h_{0}| -q|z J M - h_{0}|-r|z J M | \label{ztfren}
\end{eqnarray}

the external potential was omitted. Applying the equilibrium
condition $dF/dM = 0$ to (\ref{ztfren}) we get,

\begin{equation}
M = p\, \frac{| zJM + h_{0} |}{zJM + h_{0}} + q\, \frac{| zJM -
h_{0} |}{zJM - h_{0}} + r \frac{|zJM|}{zJM} \label{ztmagnet}
\end{equation}

Analyzing Eq. (\ref{ztmagnet}) we find a variety of solutions
because of the greater number of degrees of freedom than in the
case of the bimodal PDF \cite{asymmetric}. The solution $M = 1$ is
a stable one for $ p+r> h_{0}/zJ$, whereas for $p+r < h_{0}/zJ$
the stable one is $M = 1 - 2q$. If we consider the symmetric
trimodal PDF ($p=q=\frac{1-r}{2}$), the results found by
Sebastianes and Saxena \cite{saxena} are recovered, that is, the
former result ($M = 1$) is stable for $ \frac{1-r}{2}> h_{0}/zJ$,
whereas the latter ($M = r$) for $\frac{1-r}{2} < h_{0}/zJ$, using
the current notation. Across the line $h_{0}/zJ = p+r$ in the $(p,
h_{0})$-plane a first-order phase transition occurs between the
two ordered phases with $M = 1$ and $M = 1 - 2q$. In addition to
the aforementioned two solutions, there are more ones; the result
$M=2(p+q)-1$ is stable for $p-r>h_{0}/zJ$, $M=2p-1$ for $ p-r <
h_{0}/zJ$, $M=1-2p$ for $ r-p > h_{0}/zJ$, $M=1-2(p+q)$ for $ r-p
< h_{0}/zJ$ and $M=-1$ for $p+r+h_{0}>0/zJ$. In the first case, $M
= 1$, the condition $p > (h_{0}/(zJ))$ implies that the exchange
interaction $J$ is much stronger than randomness $h_{0}$, their
ratio is always smaller than one, thus forcing the system's spins
to order according to the first term in (\ref{rham}). The
alternative condition $p < (h_{0}/(zJ))$ implies, now, that
randomness is not any more negligible but strong enough to
influence significantly the spins enforcing a $p$-fraction of them
to point up and a $q$-fraction down, to randomly align with the
local fields, thus, practically, it dominates, so to speak, over
the first term in Eq. (\ref{rham}) so that $M = p-q+r = 1-2q$.

\vspace{-6mm}

\section{Conclusions and discussions}

\vspace{-6mm}

In the current treatment we have determined the phase diagram and
discussed some critical phenomena of the Ising model under the
influence of a trimodal random field, an extension of the bimodal
one, to allow for the existence of non magnetic particles or
vacancies in the system, for arbitrary values of the probabilities
$p$ and $q$ via the mean-field approximation. The competition
between the ordering effects and the randomness induces a rich
phase diagram. The system is strongly influenced by the random
field, which establishes a new competition favoring disorder; this
is obvious by the appearance of first-order transitions and
tricritical points, in addition to the second-order transitions,
for some values of $p$ and $q$; the tricritical point temperature
has various modes of variation as a function of $p$ and $q$. The
trimodal distribution induces reentrant behavior for the
appropriate range of $p, q$ and random field $h_{0}$. For some
values of $p$ and $q$ the system can be found either in the PM
phase or in the FM phase for low temperatures and medium and/or
high random fields; occasionally, the part of the phase diagram
allocated to the FM phase shrinks significantly. A direct
consequence of the asymmetric PDF ($p \neq q$) is the existence of
residual mean magnetization, a result of $<h_{i}>_{h} =
(p-q)h_{0}$, making the TCP non zero magnetization $M_{2}$ to be
the stable one in comparison to the zero one, $M_{1}$; however,
for $p \neq q$ (symmetric PDF) the residual mean magnetization
vanishes as well as the initially TCP non zero magnetization so
that $M_{2} \equiv M_{1}$, which is now the stable one. Both
asymmetric PDFs, bimodal and trimodal, confirm the existence of a
TCP and, nevertheless, yield similar magnetization profiles as
well as reentrance; however, the trimodal one predicts also the
existence of a second TCP. The tricritical point temperatures for
the bimodal and trimodal PDF's are independent of the random field
strength $h_{0}$; they depend only on the probability $p$ and
$p,q$, respectively.

Griffiths extended the notion of the critical point to the
so-called multicritical points, e.g., tricritical,
critical-end-point, double critical-end-point, fourth-order,
ordered critical point, etc. \cite{griffiths}; however, in order
to describe these points (except the first two) the considered
expansion of the free energy (\ref{mfafren2}) has to be extended
to higher-order terms \cite{trimodal,crok3,galambirman,milmanetal}
so that the stability criteria for such a point are satisfied, but
this is beyond the scope of the current research.

The Landau theory breaks down close to the critical point
(non-classical region) because as the transition temperature is
approached the fluctuations become important and non-classical
behavior is observed. A relative criterion, called Ginzburg
criterion, determines how close to the transition temperature the
true critical behavior is revealed, or, in other words, it governs
the validity of the Landau theory close to a critical point
\cite{ginzburg}. This criterion relies on any thermodynamic
quantity but the specific heat is usually considered for
determining the critical region around $T_{c}$ where the mean
field solution cannot describe correctly the phase transition. The
Landau theory is valid for lattice dimensionality greater than or
equal to the upper critical dimension $d_{u}=4$ in case of
presence only of thermal fluctuations. However, in the current
case the presence of random fields enhances fluctuations causing
the critical region to be wider than the one due only to the
thermal fluctuations \cite{kaufmankardar,nielsen} and the upper
critical dimension is increased by $2$ to $d_{u}=6$.

Our results indicate that on increasing the complexity of the
model system new phenomena can be revealed as in the current case
of including asymmetry in the PDF; this inclusion induces drastic
changes on the phase diagram, such as reentrance and two TCPs,
thus confirming the necessity of treating the partial
probabilities $(p,q,r)$ of the PDF in the most general aspect to
get the complete phase diagram. A similar situation appears in the
model systems in Refs. \cite{galamaharony,galam1} wherein the
considered complexity has revealed a rich variety of phase
diagrams with known and new multicritical points. The results
obtained in the current investigation by using the MFA need
further analysis; these can provide a basis for a comprehensive
analysis by more sophisticated methods as well as experimental
implementation. However, they are of no less importance, since
they show, nevertheless, the expected phenomena to be observed.

\vspace{-5mm}

\ack{The author express his gratitude to Professors A. Nihat Berker and
A. Malakis for their useful comments.

This research was supported by the Special Account for Research Grants
of the University of Athens ($E\Lambda KE$) under Grant No. 70/4/4096.}


\newpage

\begin{thebibliography}{00}

\bibitem{huiberker} K. Hui, A. Nihat Berker, Phys. Rev. Lett. 62 (1989) 2507.

\bibitem{physicstoday} D. S. Fisher, G. M. Grinstein, A. Khurana,
Phys. Today 41 (12) (1988) 56.

\bibitem{natermannvillain} T. Nattermann, J. Villain, Phase
Transitions 11 (1988) 5.

\bibitem{imryma} Y. Imry, S.-K. Ma, Phys. Rev. Lett. 35 (1975) 1399.

\bibitem{blume} M. Blume,  Phys. Rev. 141 (1966) 517.

\bibitem{capel} H. W. Capel,  Physica (Utr.) 32 (1966) 966.

\bibitem{imbrie} J. Z. Imbrie, Phys. Rev. Lett. 53 (1984) 1747.

\bibitem{aharony} A. Aharony,  Phys. Rev. B 18 (1978) 3318.

\bibitem{sneiderpytte} T. Schneider, E. Pytte, Phys. Rev. B 15 (1977) 1519.

\bibitem{fernandez}  L. A. Fern\'{a}ndez, A. Gordillo-Guerrero, V. Mart\'{\i}n-Mayor,
 J. J. Ruiz-Lorenzo, Phys. Rev. Lett. 100 (2008) 057201.

\bibitem{gofman} M. Gofman, J. Adler, A. Aharony, A. B. Harris, M. Schwartz,
Phys. Rev. Lett. 71 (1993) 2841; Phys. Rev. B 53 (1996) 6362.

\bibitem{houghton} A. Houghton, A. Khurana, F. J. Seco, Phys. Rev.
B 34 (1986) 1700.

\bibitem{galambirman2} S. Galam, J. L. Birman, Phys. Rev. B 28 (1983) 5322.

\bibitem{machta} J. Machta, M. E. J. Newman, L. B. Chayes, Phys. Rev.
E 62 (2000) 8782.

\bibitem{middleton} A. A. Middleton, D. S. Fisher, Phys. Rev. B
65 (2002) 134411.

\bibitem{fytas1} N. G. Fytas, A. Malakis, K. Eftaxias, J. Stat. Mech. (2008) P03015.

\bibitem{hernandezetal} L. Hern$\acute{a}$ndez, H. T. Diep, Phys. Rev. B 55 (1997) 14080;
L. Hern$\acute{a}$ndez, H. Ceva, Physica A 387 (2008) 2793.

\bibitem{fishaha} S. Fishman, A. Aharony, J. Phys. C: Solid State Phys. 12 (1979) L729.

\bibitem{galam3} S. Galam, Phys. Rev. B, 31 (1985) 7274.

\bibitem{andelman1} D. Andelman, Phys. Rev. B 27 (1983) 3079.

\bibitem{kaufkan} M. Kaufman, M. Kanner, Phys. Rev. B 42 (1990) 2378.

\bibitem{dgaussian} N. Crokidakis, F. D. Nobre, J. Phys.: Condens. Matter 20
(2008) 145211.

\bibitem{galamaharony} S. Galam, A. Aharony, J. Phys. C: Solid St. Phys., 13 (1980) 1065.

\bibitem{galam1} S. Galam, J. Phys. C: Solid St. Phys., 15 (1982) 529.

\bibitem{galam2} S. Galam, Europhys. Lett., 37 (1997) 615;
J. of Non-Crystalline Solids, 235-237 (1998) 570.

\bibitem{asymmetric} I. A. Hadjiagapiou, Physica A 389 (2010) 3945.

\bibitem{trimodal} M. Kaufman, P. Klunzinger, A. Khurana Phys. Rev. B 34
(1986) 4766.

\bibitem{saxena}V. K. Saxena, Phys. Rev. B 35 (1987) 2055;
R. M. Sebastianes, V. K. Saxena, \textit{ibid.} 35 (1987) 2058.

\bibitem{riegeryoung} H. Rieger, A. Peter Young, J. Phys. A: Math. Gen. 26 (1993) 5279.

\bibitem{rieger} H. Rieger, Phys. Rev. B 52 (1995) 6659.

\bibitem{hartmannyoung} A. K. Hartmann, A. P. Young, Phys. Rev. B
64 (2001) 214419.

\bibitem{nowak} U. Nowak, K. D. Usadel, J. Esser, Physica A 250 (1998) 1.

\bibitem{dukovski} I. Dukovski, J. Machta,  Phys. Rev. B 67 (2003) 014413.

\bibitem{malakisfytas} A. Malakis, N. G. Fytas,  Phys. Rev. E 73 (2006) 016109;
 Eur. Phys. J. B 50 (2006) 39.

\bibitem{belangeryoung} D. P. Belanger, A. P. Young, J. Magn. Magn. Mater. 100 (1991) 272.

\bibitem{crok2} N. Crokidakis, F. D. Nobre, Phys. Rev. E 77 (2008) 041124.

\bibitem{crok3} O. R. Salmon, N. Crokidakis, F. D. Nobre, J. Phys.: Condens.
Matter 21 (2009) 056005.

\bibitem{khurana} A. Khurana, F. J. Seco, A. Houghton, Phys. Rev. Lett. 54 (1985) 357.

\bibitem{galam} S. Galam, C. S. O. Yokoi, S. R. Salinas,  Phys. Rev B 57 (1998) 8370.

\bibitem{weizenmann} A. Weizenmann, M. Godoy, A. S. de Arruda,
D. F. de Albuquerque, N. O. Moreno, Physica B 398 (2007) 297.

\bibitem{stanley} H. Eugene Stanley, Introduction to Phase Transitions and Critical
Phenomena, Clarendon Press - Oxford, U.K., (1971), p. 43.

\bibitem{robertson} H. S. Robertson, Statistical Thermophysics, Prentice-Hall,
New Jersey, U.S.A., 1993, pp. 303, 308.

\bibitem{lawrie} I. D. Lawrie, S. Sarbach, Phase Transitions and Critical Phenomena,
eds. C. Domb and J. L. Lebowitz, Vol. 9,  Academic Press, London,
U.K., (1984).

\bibitem{hadjievans} I. Hadjiagapiou, R. Evans, Mol. Phys. 54 (1985) 383.

\bibitem{plalan} J. A. Plascak, D. P. Landau, Phys. Rev. E 67 (2003) 015103(R).

\bibitem{tsai} S.-H. Tsai, F. Wang, D. P. Landau, Brazilian Journal of Physics 36 (2006) 635;
Phys. Rev. E 75 (2007) 061108(R).

\bibitem{rowwid} J. S. Rowlinson, B. Widom, Molecular Theory of Capillarity,
Clarendon, Oxford, (1982).

\bibitem{rowswi} J. S. Rowlinson, F. L. Swinton,  Liquids and Liquid Mixtures,
Butterworths, U.K. (1982).

\bibitem{widom} B. Widom, J. Phys. Chem. 77 (1973) 2196; \textit{ibid.} 100 (1996) 13190.

\bibitem{kumar} A. Kumar, Physica A 146 (1987) 634.

\bibitem{kraska} T. Kraska, A. R. Imre, S. J. Rzoska,  J. Chem. Eng. Data 54 (2009) 1569.

\bibitem{wangkimel} Y.-L. Wang, J. D. Kimel, J. Appl. Phys. 69 (1991)
6176.

\bibitem{griffiths} R. B. Griffiths,  Phys. Rev. B 12 (1975) 345.

\bibitem{galambirman} S. Galam, J. L. Birman, J. Phys. C : Solid State Phys. 16
(1983) L1145.

\bibitem{milmanetal} F. S. Milman, P. R. Hauser, W. Figueiredo,
Phys. Rev. B 43 (1991) 13641.

\bibitem{ginzburg} V. L. Ginzburg, Fiz. Tverd. Tela. (Leningrad) 2
(1960) 2031 [Sov. Phys.-Solid State 2 (1961) 1824].

\bibitem{kaufmankardar} M. Kaufman, M. Kardar, Phys. Rev. B 31 (1985) 2913.

\bibitem{nielsen} J. Als-Nielsen, R. J. Birgeneau, Amer. J. Phys. 45 (1977) 554.

\end{thebibliography}
\end{document}